\documentstyle[12pt,epsf]{article}
\newcommand{\del}{\partial}

\newcommand{\dis}{\displaystyle}
\newcommand{\ba}{\begin{eqnarray}}
\newcommand{\ea}{\end{eqnarray}}
\newcommand{\non}{\nonumber\\}
\newcommand{\nom}{\nonumber}
\newcommand{\vpi}{\varpi}
\newcommand{\tpi}{\tilde{\varpi}}
\newcommand{\kae}{\mbox{K\"ahler}}
\newcommand{\kei}{\mbox{\Large\mbox{$\kappa$}}}
\begin{document}

%%%%%%%%%%%%%%%%%%% titlepage %%%%%%%%%%%%%%%%%%%%%
\begin{titlepage}
\nopagebreak
%%%%%%%%%%%%%%%%%%%%%%%%%%%%%%%%%%%%%%%%%%%%%%%%%%%%
\begin{flushright}
March 2000\hfill 
%KUCP-\\
hep-th/0003164
\end{flushright}

\vfill
\begin{center}
{\LARGE K\"ahler Potential of Moduli Space}\\

~

{\LARGE of Calabi-Yau $d$-fold}\\

~

{\LARGE  embedded in $CP^{d+1}$}

\vskip 12mm

{\large Katsuyuki~Sugiyama${}^{\dag}$}

\vskip 10mm
{\sl Department of Fundamental Sciences}\\
{\sl Faculty of Integrated Human Studies, Kyoto University}\\
{\sl Yoshida-Nihon-Matsu cho, Sakyo-ku, Kyoto 606-8501, Japan}\\
\vspace{1cm}

     \begin{large} ABSTRACT \end{large}
\par
\end{center}
%\vfill
\begin{quote}
 \begin{normalsize}

%%%%%%%%%%%%%%%%%%%%%%%%%%%%%%%%%%%%%%%%%%%%%%%%
%\begin{abstract}
We study a {\kae} potential $K$ of a one parameter family of Calabi-Yau 
$d$-fold embedded in $CP^{d+1}$.
By comparing results of the topological B-model and 
the data of the CFT calculation at Gepner point, 
the $K$ is determined unambiguously.
It has a moduli parameter $\psi$ that describes a 
deformation of the CFT by a marginal operator.
Also the metric, curvature and hermitian two-point functions 
in the neighborhood of the Gepner point 
are analyzed. 
We use a recipe of $tt^{\ast}$ fusion and 
develop a method to determine the $K$ from the point of view of 
topological sigma model. It is not restricted to this specific model
and can be applied to other Calabi-Yau cases.
%\end{abstract}

\end{normalsize}
\end{quote}

%\mbox{}\\
%PACS codes; 11.25.-w, 11.25.Mj, 11.25.Sq, 02.\\ 
%PACS codes; 11.25.-w, 11.25.Mj, 11.25.Sq, 11.25.Hf, 11.10.Kk, 02.\\ 
%Keywords; \\
\mbox{}\hspace{21mm}
\vfill 
\noindent
\rule{7cm}{0.5pt}\par
\vskip 1mm
{\small \noindent ${}^{\dag}$  
E-mail :\tt{} sugiyama@phys.h.kyoto-u.ac.jp}
\end{titlepage}
\vfill

%%%%%%%%%%%%%%%%%%%%%%%%%%%%%%%%%%%%%%%%%%%%%%%%%%%%
%   Introduction
%%%%%%%%%%%%%%%%%%%%%%%%%%%%%%%%%%%%%%%%%%%%%%%%%%%%
\section{Introduction}
%\cleqn

There are great progresses in studying worldsheet instanton
corrections by the discovery of the mirror symmetry \cite{GP},\cite{CDGP}.
When we study topological sigma model \cite{W}, physical observables are 
constructed by combining couplings, correlation functions and metrics
of operators. Typical constituent blocks are three-point and two-point 
functions (metrics). Their behaviors are controlled by moduli
parameters and these observables contain information about 
moduli space of the sigma model \cite{GMP}-\cite{KS6}. 
In this paper, we investigate a
{\kae} potential of the B-model moduli space from the point of view of 
the topological sigma model together with results by CFT at Gepner
point.

\section{Calabi-Yau $d$-fold}
We concentrate on a one-parameter family of
Calabi-Yau $d$-fold $M$ realized as a 
zero locus of a hypersurface in a projective space $CP^{d+1}$
\ba
&&W\,;\,\widehat{\{p=0\}/{\bf Z}_N^{\otimes (N-1)}}\,,\non
&&p=X^N_1+X^N_2+\cdots +X^N_N-N\psi X_1X_2\cdots X_N=0\,,\nom
\ea
where we introduce a number $N=d+2$.
Hodge numbers $h^{p,q}$ of this $d$ fold are calculated as \cite{Dais} 
\ba
&&h^{p,q}=\delta_{p+q.d}\,,\,\,\,\,(0\leq p\leq d\,,
0\leq q\leq d\,,p\neq q)\,,\non
&&h^{p,p}=\delta_{2p,d}+\sum^{p}_{m=0}
(-1)^m\left(\matrix{d+2\cr m}\right)\cdot 
\left(\matrix{(p+1-m)(d+1)+p\cr d+1} \right)\,,\nom
\ea
and deformation of complex structure is 
parametrized by a complex parameter $\psi$.

In order to describe the complex structure 
moduli space near Gepner point
($\psi\sim 0$), we uses a basis of periods
\ba
&&\tpi_k(\psi)
=\left[\Gamma\!\left(\frac{k}{N}\right)\right]^N
\frac{(N\psi)^{k}}{\Gamma (k)}\cdot G_k(\psi)\,,\non
&&G_{k}:=1+
\sum_{n=1}^{\infty}
\left[\frac{\Gamma\!\left(\frac{k}{N}+n\right)}
{\Gamma\!\left(\frac{k}{N}\right)}\right]^N\frac{\Gamma (k)}{\Gamma (Nn+k)}
(N\psi)^{Nn}\,.\nom
\ea

They behave around $\psi \sim 0$
\ba
\tpi_k(\psi)=\left[\Gamma\!\left(\frac{k}{N}\right)\right]^N
\frac{(N\psi)^{k}}{\Gamma (k)}\cdot \left[1+{\cal O}(\psi^N)\right]\,.\nom
\ea
This set is a projective coordinate of the B-model moduli space.
Also we use a canonical basis $\{\vpi^{(0)}_k\}$ 
of the coordinates at the Gepner point as 
\ba
\tpi_{k}
=\vpi^{(0)}_k\cdot \left[\Gamma\!\left(\frac{k}{N}\right)\right]^N\,,\nom
\ea
and take ratios to construct normalized coordinates
\ba
\omega_{k-1}:=\vpi^{(0)}_k/\vpi^{(0)}_1 
=\frac{(N\psi)^{k-1}}{(k-1)!}\cdot \frac{G_k}{G_1}
\,\,\,\,\,\,\,(k=1,2,\cdots ,N-1)\,.\nom
\ea
Some part of the cohomology classes of $H^d(W)$ is associated with the 
complex structure deformation
with Hodge numbers $h^{\ell ,d-\ell}=1$ $(\ell =0,1,2,\cdots ,d)$.
Then associated operators  
are represented as
\ba
{\cal O}^{(k)}:=(X_1X_2\cdots X_N)^k\,\,\,\,(k=0,1,2,\cdots ,N-2)\,,\nom
\ea
and each  ${\cal O}^{(k)}$ is related to a canonical period $\omega_{k}$.
Also a coupling $\tilde{\psi}:=N\psi$
is associated with an operator 
${\cal O}^{(1)}=X_1X_2\cdots X_N$.
When we  choose a canonical set of periods, these operators are 
normalized appropriately and 
fusion couplings ${\mbox{\Large \mbox{$\kappa$}}_{\ell} }$ 
are evaluated at $\psi =0$
\ba
&&\vec{\omega}:={}^t(\matrix{\omega_0 & \omega_1 &\cdots &
  \omega_{N-2}})\,,\non
&&{\bf K}:=\left(\matrix{0 & \kei_0 & & & & \cr & 0 & \kei_1 & & & \cr
& & 0 & \kei_2 & &\cr   
& & & \ddots & \ddots & \cr & & & & 0 & \kei_{d-1} 
\cr 0 & 0 & 0 & \cdots & 0 & 0}\right)
\,\non
&&\del_{\tilde{\psi}}\vec{\omega}={\bf K}\vec{\omega}\,,\non
&&\leftrightarrow {\cal O}^{(1)}\cdot {\cal O}^{(\ell)}=
  \mbox{\Large \mbox{$\kappa$}}_{\ell} 
{\cal O}^{(\ell +1)}\,\,\,\,(\ell =0,1,\cdots ,d-1)\,.\non
&&\qquad \mbox{\Large\mbox{$\kappa$}}_{0}=1\,,\,\,\,
\mbox{\Large\mbox{$\kappa$}}_{1}=
\del\omega_1=1+{\cal  O}(\psi^N)\,,\,\,\,\non
&&\qquad \mbox{\Large\mbox{$\kappa$}}_{\ell}=
\del \frac{1}{ \kei_{\ell -1}}\del \frac{1}{ \kei_{\ell -2}}
\del \cdots \del \frac{1}{ \kei_{0}}\del \omega_{\ell}
=1+{\cal O}(\psi^N)\,,\,\,\,\,\,
(\ell \geq 2)\,.\nom
\ea
In this canonical set of periods, 
the three-point couplings are 
normalized to units up to terms with order ${\cal O}(\psi^N)$.

\section{{\kae} potential}
Next let us consider a 
{\kae} potential $K$ of the moduli space.
We can construct $K$ as a quadratic form of periods
\ba
e^{-K}=i^{d^2}\int_M\Omega\wedge\bar{\Omega}=
\sum_{m,n=1}^{N-1}I_{m,n}\tpi_{m}^\dagger \tpi_{n}\,.\nom
\ea
The matrix $I=\{I_{m,n}\}$ is determined by properties of intersection 
numbers of homology cycles and does not change under 
any infinitesimal deformation of continuous moduli parameters.
That is, the $I_{m,n}$s cannot depend on any moduli parameters and turns
out to be constant numbers.

On the other hand, global structures of the moduli space are
encoded in monodromies and it is important to
 see property of $K$ under a global monodromy transformation.
Our basis $\{\tpi_k\}$ 
diagonalizes a cyclic ${\bf Z}_N$ monodromy $\psi\rightarrow
\alpha\psi$ ($\alpha =e^{2\pi i/N}$)
at $\psi =0$
\ba
&&\tpi_k(\alpha\psi)=\alpha^k \tpi_k(\psi)\,\,\,\,\,(k=1,2,\cdots ,N-1)\,.\non
&&\alpha =e^{2\pi i/N}\,.\nom
\ea
But this transformation induces a change of the $I_{m,n}$
\ba
I_{m,n}\rightarrow I_{m,n}\cdot \alpha^{-m+n}\,\,\,(m,n=1,2,\cdots
,N-1)\,.\nom
\ea
Because the {\kae} potential is physical 
quantity and should not depend on monodromies, it is 
invariant under the transformation.  
Only invariant parts of this transformation are diagonal ones
$I_{m,m}$ $(m=1,2,\cdots ,N-1)$ and
we find that
the matrix $I$ is diagonal one:
\ba
e^{-K}=\sum_{k=1}^{N-1}I_{k}\tpi^\dagger_k\tpi_k\,.\label{KO}
\ea

\section{Determination of $I_k$}

Next, in order to determine the {\kae} potential, 
all we have to do is to fix the diagonal matrix. 
We use a method of the $tt^{\ast}$ fusion in fixing the 
$I_k$ ($k=1,2,\cdots ,N-1$).
First note that the {\kae} potential is related to 
a moduli space metric $g_{\tilde{\psi}\bar{\tilde{\psi}}}$
\ba
&&\del\bar{\del}K=g_{\tilde{\psi}\bar{\tilde{\psi}}}\,,\,\,\,\,\,
\del =\frac{\del}{\del\tilde{\psi}}\,,\,\,
\bar{\del} =\frac{\del}{\del\bar{\tilde{\psi}}}\,.\,\,\nom
\ea
When we introduce a set of hermitian two-point functions 
\footnote{These two-point functions are different from topological metrics 
$\mbox{\large\mbox{$\eta$}}_{\ell m}
=\langle{\cal O}^{(\ell)}{\cal O}^{(m)}\rangle =N\delta_{\ell +m,d}$.}
\ba
\langle\bar{{\cal O}}^{(\ell)}|{\cal O}^{(\ell)}\rangle=
e^{q_{\ell}}\,\,\,\,\,\,\,(0\leq \ell \leq N-2)\label{zamo}\,,
\ea
the {\kae} potential and Zamolodchikov metric 
$g_{\tilde{\psi}\bar{\tilde{\psi}}}$
are expressed as
\ba
&&e^{q_0}=e^{-K}\,,\,\,\,
g_{\tilde{\psi}\bar{\tilde{\psi}}}=e^{q_1-q_0}\,,\non
&&-\del\bar{\del}q_0=e^{q_1-q_0}\,.\label{kae}
\ea
The equation Eq.(\ref{kae}) is a part of the $t t^{\ast}$-fusion
equation \cite{tt,KS5} of the Calabi-Yau $d$-fold 
with fusion couplings $\mbox{\Large\mbox{$\kappa$}}_n$
\ba
&&\del\bar{\del}q_0+|\mbox{\Large\mbox{$\kappa$}}_0|^2
e^{q_1-q_0}=0\,,\non
&&\del\bar{\del}q_{\ell}+|\mbox{\Large\mbox{$\kappa$}}_{\ell}|^2
e^{q_{\ell +1}-q_{\ell}}-|\mbox{\Large\mbox{$\kappa$}}_{\ell -1}|^2
e^{q_{\ell}-q_{\ell -1}}
=0\,\,\,(1\leq \ell\leq d-1)\,,\non
&&\del\bar{\del}q_d-|\mbox{\Large\mbox{$\kappa$}}_{d-1}|^2
e^{q_d-q_{d-1}}=0\,.\label{toda1}
\ea
This set of equations Eq.(\ref{toda1}) represents an $A_{d}$ type Toda system.
By introducing new variables
\ba
&&\tilde{q}_0=q_0\,,\non
&&\tilde{q}_{\ell}=q_{\ell}+\sum_{n=0}^{\ell -1}\log 
|\mbox{\Large{\mbox{$\kappa$}}}_n|^2\,\,\,\,(\ell \geq 1)\,,\nom
\ea
and noting relations $\del\bar{\del}\tilde{q}_{\ell}=\del\bar{\del}q_{\ell}$,
we reexpress the Toda system into a formula
\ba
&&U_{\ell}=(\ell +1)U_0+\sum^{\ell -1}_{n=0}(\ell -n)D\log
(-U_n)\,\,\,\,\,(1\leq \ell \leq d-1)\,,\non
&&U_0=Dq_0=D\log \left[
1+\sum^{N-1}_{\ell =2}\frac{I_{\ell}}{I_1}
\left|\frac{\tpi_{\ell}}{\tpi_1}\right|^2\right]\,,\non
&&U_d=0\,,\non
&&U_n=\sum^{n}_{m=0}Dq_{m} \,\,\,\,\,(1\leq n\leq d)\,,\non
&&D:=\del\bar{\del}\,,\non
&&|\mbox{\Large\mbox{$\kappa$}}_n|^2\cdot e^{q_{n+1}-q_n}=-U_n\,\,\,
\,\,\,\,(0\leq n\leq d-1)\,\label{toda2}
\ea
The number of diagonal components $I_k$ in the {\kae} potential  is
$(d+1)$ and the above Toda system 
with (d+1) equations gives us consistency conditions
of the $I_k$s. 
Now recall that the components of the matrix $I$ are numerical
constants. We may truncate terms of the periods
$\tpi_k$s in the series expansions up to order ${\cal O}(\psi^{N})$
for the purpose of determination of $I$
\ba
&&\tpi_k(\psi)=f_k \cdot (N\psi)^{k}
\left[1+{\cal   O}(\psi^N)\right]\,\,\,\,\,
(k=1,2,\cdots ,N-1)\,,
\non
&&U_0=D\log\left[1+\sum^{N-1}_{\ell =2}a_{\ell }
(\tilde{\psi}\bar{\tilde{\psi}})^{\ell   -1}+\cdots 
\right]\,.\non
&&f_k:=
\left[\Gamma\!\left(\frac{k}{N}\right)\right]^N\frac{1}{\Gamma (k)}\,,
\,\,\,a_n:=\frac{I_{n}}{I_1}
\left|\frac{f_n}{f_1}\right|^2\,.\label{an}
\ea
We calculate the $U_n$ concretely for $n\leq 10$, 
and propose a conjecture for the $\{U_n\}$s at the Gepner point $\psi =0$
\ba
&&U_0=a_2\,,\,\,U_d=0\,,\non
&&U_n=(n+1)^2\frac{a_{n+2}}{a_{n+1}}\,\,\,\,\,\,(1\leq n\leq d-1)\,.\nom
\ea
When we use definitions of $a_n$ and $f_n$, the $U_n$s are expressed as
\ba
&&U_0=a_2=\frac{I_2}{I_1}\left[
\frac{\Gamma\!\left(\frac{2}{N}\right)}{\Gamma\!\left(\frac{1}{N}\right)}
\right]^{2N}\,,\non
&&U_n=(n+1)^2\frac{a_{n+2}}{a_{n+1}}=
\frac{I_{n+2}}{I_{n+1}}\left[
\frac{\Gamma\!\left(\frac{n+2}{N}\right)}{\Gamma\!\left(\frac{n+1}{N}\right)}
\right]^{2N}\,\,\,\,(1\leq n\leq d-1)\,.\nom
\ea
Also we can obtain expressions of hermitian two-point functions
by remarking that the three-point couplings become constants 
$\mbox{\Large\mbox{$\kappa$}}_n=1$ $(n=0,1,\cdots ,d)$ at the $\psi =0$
\ba
e^{q_{n+1}-q_n}=
-\frac{I_{n+2}}{I_{n+1}}
\left[\frac{\Gamma\!\left(\frac{n+2}{N}\right)}
{\Gamma\!\left(\frac{n+1}{N}\right)}\right]^{2N}\,\,\,(0\leq n\leq d-1)\,.\nom
\ea
Equivalently, normalized two-point functions are represented as
\ba
\frac{\langle\bar{\cal O}^{(m)}|{\cal O}^{(m)}\rangle}
{\langle\bar{\cal O}^{(0)}|{\cal O}^{(0)}\rangle}=
e^{q_m-q_0}=(-1)^m\cdot \frac{I_{m+1}}{I_1}
\left[\frac{\Gamma\!\left(\frac{m+1}{N}\right)}
{\Gamma\!\left(\frac{1}{N}\right)}\right]^{2N}\,\,\,\,\,\,(0\leq 
m\leq d)\,.\label{result}
\ea
Now let us compare these results with those of CFT calculations
\cite{CFT,tt} for minimal models associated with the $d$-fold $M$
\ba
&&e^{q_n}=\frac{1}{N^N}\left[\frac{\Gamma\!\left(\frac{n+1}{N}\right)}
{\Gamma\!\left(1-\frac{n+1}{N}\right)}\right]^N\,\,\,\,
(n=0,1,2,\cdots ,N-2)\,,\non
&&\rightarrow e^{q_n-q_0}=
\left[\frac{\Gamma\!\left(\frac{n+1}{N}\right)
\Gamma\!\left(1-\frac{1}{N}\right)}
{\Gamma\!\left(\frac{1}{N}\right)\Gamma\!\left(1-\frac{n+1}{N}\right)}
\right]^N\,.\label{cft}
\ea
By comparing two results Eq.(\ref{result}) and Eq.(\ref{cft}),
we can obtain the components of the matrix $I$ up to one numerical
constant $c$
\ba
I_m=c\cdot (-1)^m \left(\sin\frac{\pi m}{N}\right)^N\,\,\,\,
(m=1,2,\cdots ,N-1)\,.\nom
\ea
In this case, the $e^{-K}$ in Eq.(\ref{KO}) is given as
\ba
&&e^{-K}= \sum_{m=1}^{N-1}
c\cdot (-1)^m \pi^N\left[
\frac{\Gamma\!\left(\frac{m}{N}\right)}{\Gamma\!\left(1-\frac{m}{N}\right)}
\right]^N\times \frac{(N^2\psi\bar{\psi})^m}{[\Gamma\!(m)]^2}\cdot 
|G_m(\psi)|^2
\,.\nom
\ea
Because the {\kae} potential is not a function but a section of a line
bundle, there is an arbitrariness of multiplication of arbitrary 
(anti-)holomorphic functions.
We choose the normalization factor as
\ba
c=\frac{-1}{\pi^N\cdot N^{N+2}}\,,\nom
\ea
then the {\kae} potential is written as
\ba
e^{-K}=(\psi\bar{\psi})\cdot 
\sum^{N-1}_{m=1}(-1)^{m-1}\left[\frac{1}{N}
\frac{\Gamma\!\left(\frac{m}{N}\right)}
{\Gamma\!\left(1-\frac{m}{N}\right)}\right]^N
\frac{(N^2\psi\bar{\psi})^{m-1}}{[\Gamma\!(m)]^2}|G_m(\psi)|^2
\,.\label{potential}
\ea
In order to confirm the validity of this convention, we restrict
ourselves to the 3-fold $(N=5)$ case and consider a $3$-point function
${\kei}$ of the operator ${\cal O}^{(1)}$ in the B-model
\ba
\kei :=\kei_{\psi\psi\psi}=
\frac{1}{5^3}\cdot\frac{5\psi^2}{1-\psi^5}\,.\label{3-point}
\ea
(In considering the $tt^{\ast}$ equation, we use normalized periods 
$\omega_k$ by taking ratios. In that case, the $e^{-K}$ in
Eq.(\ref{potential}) and the three-point function Eq.(\ref{3-point}) are
 divided by a factor $\sim |\psi|^2$. )
This coupling is a section of holomorphic line bundle and changes
with a normalization of the {\kae} potential.
But there is an invariant $3$-point coupling 
\ba
{(g_{\psi\bar{\psi}})}^{-3/2}\cdot e^K \cdot |\kei|\,,\nom
\ea
and its value is evaluated at the $\psi =0$ in our normalization
\ba
{(g_{\psi\bar{\psi}})}^{-3/2}\cdot e^K \cdot |\kei|=
\left[\frac{\Gamma\!\left(\frac{3}{5}\right)}
{\Gamma\!\left(\frac{2}{5}\right)}\right]^{15/2}
\cdot \left[\frac{\Gamma\!\left(\frac{1}{5}\right)}
{\Gamma\!\left(\frac{4}{5}\right)}\right]^{5/2}=
1.5553189899632389725\cdots \,.\nom
\ea
This result coincides with that of the calculation of CFT \cite{CFT,CDGP}.

\section{Metric and Curvature}

Now we have an exact formula of the $K$ with a moduli parameter 
$\psi $ and can
evaluate the corrections in the physical observables
by the marginal deformation of the CFT.
For simplicity, we will consider leading corrections 
of the {\kae} potential, metric and scalar curvature in the moduli space.
When we define a function $A_m$
\ba
A_m=(-N^2)^{m-1}\cdot\frac{1}{[\Gamma\!(m)]^2}
\left[\frac{1}{N}\frac{\Gamma\!\left(\frac{m}{N}\right)}
{\Gamma\!\left(1-\frac{m}{N}\right)}\right]^N\,\,\,\,\,\,\,
(1\leq m\leq N-1)\,,\nom
\ea
we can obtain these formulae
\ba
&&\frac{A_m}{A_n}=(-N^2)^{m-n}\left(\frac{\Gamma\!(n)}{\Gamma\!(m)}\right)^2
\left[\frac{\Gamma\!\left(\frac{m}{N}\right)
\Gamma\!\left(1-\frac{n}{N}\right)}{\Gamma\!\left(1-\frac{m}{N}\right)
\Gamma\!\left(\frac{n}{N}\right)}\right]^N\,\,\,\,\,(m,n\leq N-1)\,,\non
&&e^{-K}=
\left[\frac{1}{N}\frac{\Gamma\!\left(\frac{1}{N}\right)}
{\Gamma\!\left(1-\frac{1}{N}\right)}\right]^N\cdot
(\psi\bar{\psi})
\times
\left[1-N^2\cdot \left[\frac{\Gamma\!\left(\frac{2}{N}\right)
\Gamma\!\left(1-\frac{1}{N}\right)}
{\Gamma\!\left(1-\frac{2}{N}\right)\Gamma\!\left(\frac{1}{N}\right)}\right]^N
(\psi\bar{\psi})+\cdots \right]\,\,\,\,(N\geq 3)\,,\non
&&g_{\psi\bar{\psi}}=
N^2\cdot \left[\frac{\Gamma\!\left(\frac{2}{N}\right)
\Gamma\!\left(1-\frac{1}{N}\right)}
{\Gamma\!\left(1-\frac{2}{N}\right)\Gamma\!\left(\frac{1}{N}\right)}\right]^N
+(\psi\bar{\psi})
\left[2\cdot\left(\frac{A_2}{A_1}\right)^2-4\left(\frac{A_3}{A_1}\right)\right]
+\cdots\,\,\,\,\,(N\geq 4)\,,\non
&&R=
-4+2\cdot \left[\frac{\Gamma\!\left(\frac{1}{N}\right)
\Gamma\!\left(\frac{3}{N}\right)}
{\Gamma\!\left(1-\frac{1}{N}\right)\Gamma\!\left(1-\frac{3}{N}\right)}\right]^N
\left[\frac{\Gamma\!\left(1-\frac{2}{N}\right)}
{\Gamma\!\left(\frac{2}{N}\right)}\right]^{2N}\non
&&\qquad +(\psi\bar{\psi})\left[
24\left(\frac{A_3}{A_2}\right)-96 \left(\frac{A_1}{A_2}\right)
\left(\frac{A_3}{A_2}\right)^2+72\left(\frac{A_1}{A_2}\right)
\left(\frac{A_4}{A_2}\right)
\right]+\cdots \,\,\,\,\,(N\geq 5)\,.\nom
\ea
The Ricci tensor of the moduli space is represented as 
$\dis R_{\psi\bar{\psi}}=\frac{1}{2}g_{\psi\bar{\psi}}R$.
For the torus ($N=3$) case, its metric is a standard one of the upper-half
plane and the scalar curvature is constant
$R=-4$ except for points $\psi =e^{2\pi i\ell/3}$ ($\ell =0,1,2$).
Also the K3 ($N=4$) metric is given by the above formula and related 
scalar curvature is a negative constant number $R=-2$ except for 
$\psi =e^{\pi i\ell /2}$ ($\ell =0,1,2,3$).

\begin{figure}[htbp]
%\begin{center}
%\psbox[height=7cm]{****}
%\end{center}
\epsfxsize=8cm
 \centerline{\epsfbox{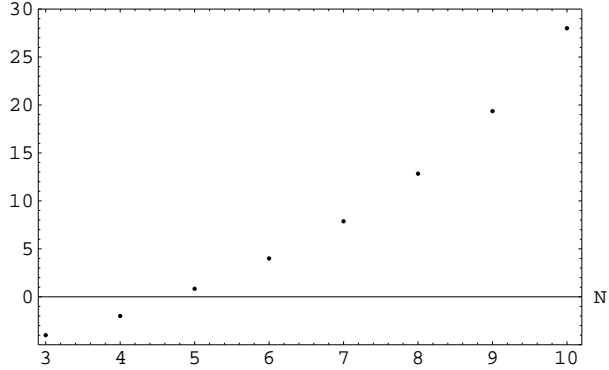}}
\epsfxsize=8cm
  \caption{Leading part $R_0$ of Scalar Curvature $R$. 
This part depends on the dimension $N=d+2$ of the Calabi-Yau
$d$-fold. It increases monotonically with the $d$.
For the torus ($N=3$) and K3 ($N=4$) cases, associated curvatures are 
negative. But the other cases have positive curvatures at $\psi =0$.}
\label{R0} 
\end{figure}

\begin{figure}[htbp]
%\begin{center}
%\psbox[height=7cm]{****}
%\end{center}
\epsfxsize=8cm
 \centerline{\epsfbox{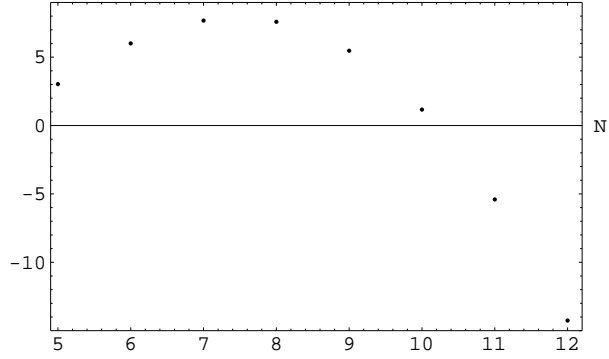}}
\epsfxsize=8cm
  \caption{Subleading part $R_1$ of Scalar Curvature $R$.
This part is a coefficient of $\psi\bar{\psi}$ in the $R$ and depends
on the dimension of the CY $d$-fold. For the $5\leq N\leq 10$ cases, 
its value is positive. But the $R_1$ is negative for 
the other cases $N\geq 11$. }
\label{R1} 
\end{figure}

We plot this scalar curvature at the $\psi =0$ in Fig.\ref{R0}.
It increases monotonically with the dimension $d$.
But the derivative with respect to $\del_{\psi}\del_{\bar{\psi}}$ is 
negative for $N\geq 11$ cases at $\psi =0$ as shown in Fig.\ref{R1}.

Let us return to 
the hermitian two-point functions. They are represented as some combinations 
of these $K$, $g_{\psi\bar{\psi}}$ and $R$

\ba
&&e^{\tilde{q}_0}=e^{-K}\,,\,\,\,
e^{\tilde{q}_1-\tilde{q}_0}=\frac{1}{N^2}
g_{\psi\bar{\psi}}\,,\,\,\,
e^{\tilde{q}_2-\tilde{q}_1}=
\frac{1}{N^2}g_{\psi\bar{\psi}}
\left(\frac{R}{2}+2\right)\,,\non
&&e^{\tilde{q}_3-\tilde{q}_2}=\frac{1}{N^2}
g_{\psi\bar{\psi}}
\left[3\left(\frac{R}{2}+1\right)
-g^{\psi\bar{\psi}}\del_{\psi}\bar{\del}_{\bar{\psi}}
\log\left(\frac{R}{2}+2\right)\right]\,,\non
&&e^{\tilde{q}_4-\tilde{q}_3}=\frac{1}{N^2}
g_{\psi\bar{\psi}}
\Biggl[4+3R-2 g^{\psi\bar{\psi}}\del_{\psi}\bar{\del}_{\bar{\psi}}
\log\left(\frac{R}{2}+2\right)\non
&&\qquad \qquad 
-g^{\psi\bar{\psi}}\del_{\psi}\bar{\del}_{\bar{\psi}}
\log\Bigl[
3\left(\frac{R}{2}+1\right)
-g^{\psi\bar{\psi}}
\del_{\psi}\bar{\del}_{\bar{\psi}}
\log\left(\frac{R}{2}+2\right)\Bigr]\Biggr]\,,\non
&&\qquad \cdots\,.\nom
\ea
Now we know the formula of the $K$, $R$, and $g_{\psi\bar{\psi}}$
and evaluate moduli dependences of these correlators.

In this paper we develop a method to 
determine the {\kae} potential unambiguously 
by comparing the result of CFT with that of topological sigma model. 
We calculate the metric and curvature
in the neighborhood of 
the Gepner point, which have dependences of moduli parameter
$\psi$. The result represents a marginal deformation of the CFT.
But the formula of the $K$ is exact and we can study it at all points in
the moduli space by analytic continuation. 
The method we developed here is not restricted to the specific model
and can be applied to any other Calabi-Yau spaces.

\section*{Acknowledgement}
This work is supported by the Grant-in-Aid for 
Scientific Research from the Ministry of Education, Science and
Culture 10740117.

\end{document}